\documentclass[twocolumn,aps,prl,showpacs,superscriptaddress,unsortedaddress]{revtex4}
\usepackage{graphicx}
\usepackage{epsf}
\usepackage{color}
\newcommand{\etal}{{\it et al.}}

\begin{document}
\title{Spectroscopic evidence for preformed Cooper pairs in the pseudogap phase of cuprates}
\author{M. Shi}
\affiliation{Swiss Light Source, Paul
Scherrer Institute, CH-5232 Villigen PSI, Switzerland}
\author{A. Bendounan}
\affiliation{Paul Scherrer Institute, ETH Zurich and EPF Lausanne,
5232 Villigen PSI, Switzerland}
\author{E. Razzoli}
\affiliation{Swiss Light Source, Paul Scherrer Institute, CH-5232
Villigen PSI, Switzerland}
\author{S. Rosenkranz}
\affiliation{Materials Science Division, Argonne National
Laboratory, Argonne, IL 60439 USA}
\author{M. R. Norman}
\affiliation{Materials Science Division, Argonne National
Laboratory, Argonne, IL 60439 USA}
\author{J. C. Campuzano}
\affiliation{Department of Physics, University of Illinois at
Chicago, Chicago, IL 60607 USA}
\affiliation{Materials Science
Division, Argonne National Laboratory, Argonne, IL 60439 USA}
\author{J. Chang}
\affiliation{Paul Scherrer Institute, ETH Zurich and EPF Lausanne,
5232 Villigen PSI, Switzerland}
\author{M. M\aa{}nsson}
\affiliation{Laboratory for Neutron Scattering, ETH Zurich and Paul
Scherrer Institute, CH-5232 Villigen PSI, Switzerland}
\affiliation{Materials Physics, Royal Institute of Technology KTH,
S-164 40 Kista, Sweden}
\author{Y. Sassa}
\affiliation{Paul Scherrer Institute, ETH Zurich and EPF Lausanne,
5232 Villigen PSI, Switzerland}
\author{T. Claesson}
\affiliation{Materials Physics, Royal Institute of Technology KTH,
S-164 40 Kista, Sweden}
\author{O. Tjernberg}
\affiliation{Materials Physics, Royal Institute of Technology KTH,
S-164 40 Kista, Sweden}
\author{L. Patthey}
\affiliation{Swiss Light Source, Paul Scherrer Institute, CH-5232
Villigen PSI, Switzerland}
\author{N. Momono}
\affiliation{Department of Physics, Hokkaido University Ð Sapporo
060-0810, Japan}
\author{M. Oda}
\affiliation{Department of Physics, Hokkaido University Ð Sapporo
060-0810, Japan}
\author{M. Ido}
\affiliation{Department of Physics, Hokkaido University Ð Sapporo
060-0810, Japan}
\author{S. Guerrero}
\affiliation{Condensed Matter Theory Group, Paul Scherrer Institute,
CH-5232 Villigen PSI, Switzerland}
\author{C. Mudry}
\affiliation{Condensed Matter Theory Group, Paul Scherrer Institute,
CH-5232 Villigen PSI, Switzerland}
\author{J. Mesot}
\affiliation{Paul Scherrer Institute, ETH Zurich and EPF Lausanne,
5232 Villigen PSI, Switzerland}

\begin{abstract}
Angle-resolved photoemission on underdoped
La$_{1.895}$Sr$_{0.105}$CuO$_4$ reveals that in the pseudogap phase,
the dispersion has two branches located above and below
the Fermi level with a minimum at the Fermi momentum. This is
characteristic of the Bogoliubov dispersion in the superconducting
state.  We also observe
that the superconducting and pseudogaps have the same
d-wave form with the same amplitude. Our observations provide
direct evidence for
preformed Cooper pairs, implying that the pseudogap phase is a precursor to
superconductivity.
\end{abstract}

\date{\today}
\maketitle

In conventional
superconductors, the energy gap decreases with increasing temperature and
vanishes at $T_c$. For the high temperature cuprates, the situation is more 
complicated. In the underdoped
region, an energy gap, known as the pseudogap, persists above $T_c$,
its maximum amplitude remaining unchanged before the gap fills in at
a temperature, $T^*$~\cite{Loeser,Ding}.  Both $T^*$ and the
pseudogap  increase with reduced doping, whereas $T_c$ decreases. In
spite of much effort, there is no consensus on the origin of the
pseudogap~\cite{NPK}. One idea is that the energy gap above
$T_c$ is indicative of preformed Cooper
pairs~\cite{Randeria,Emery} that only become phase coherent
below $T_{c}$. A competing view is that the pseudogap results from
some other order which competes with superconductivity.  In this
picture, the superconducting gap only exists along gapless Fermi
arcs~\cite{Tanaka,Kondo,Terashima,Lee}, and as in
conventional superconductors, it closes at $T_c$.
Using angle-resolved photoemission spectroscopy (ARPES),
we show the existence of a Bogoliubov-like
dispersion in the pseudogap phase of the underdoped cuprate
La$_{1.895}$Sr$_{0.105}$CuO$_4$. This provides direct evidence that
the pseudogap is a signature of preformed Cooper pairs above $T_c$.

ARPES experiments were carried out at the Surface and Interface
Spectroscopy beamline at the Swiss Light Source on single
crystals of underdoped La$_{1.895}$Sr$_{0.105}$CuO$_4$ (LSCO) with
$T_c$ = 30K and overdoped Bi$_2$Sr$_2$CaCu$_2$O$_{8+\delta}$
(Bi2212) with $T_c$ = 65 K, grown using the travel solvent floating
method. Circularly polarized light with {\it h}$\nu$ = 55 eV and
linearly polarized light with {\it h}$\nu$ = 21.2 eV were used for
LSCO and Bi2212, respectively. The energy and angle resolutions were
17 meV and 0.1 - 0.15$^\circ$. The Fermi level was
determined by recording the photoemission spectra from
polycrystalline copper on the sample holder. For LSCO, the samples
were cleaved in situ by using a specially designed
cleaver~\cite{Mansson}.

\begin{figure*}
\includegraphics
[width=6.in]
 {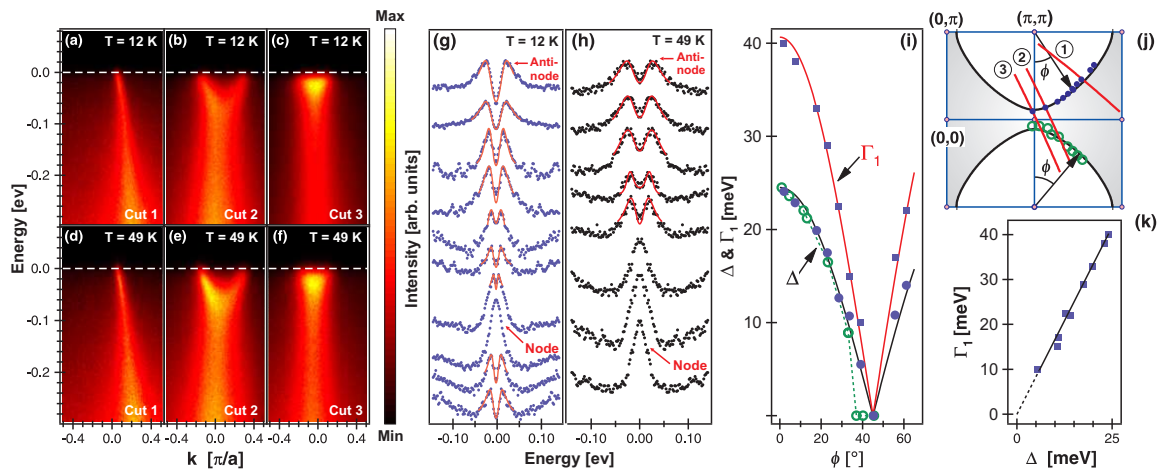}
 \caption{(Color online) ARPES spectra for underdoped La$_{1.895}$Sr$_{0.105}$CuO$_4$
($T_c$ = 30K). (a)-(c): ARPES intensities at $T$ = 12K along
momentum cuts that cross the Fermi surface at the node, between the
node and the anti-node, and at the anti-node. The corresponding cuts
are indicated in (j). (d)-(f): the same as (a)-(c), but the spectra
are acquired at 49K. (g)-(h): the symmetrized spectra for various
$k_F$ from the anti-node (top) to the node (bottom) at 12K and 49K,
respectively. The solid curves are fits. In (g), the node is the
third spectrum from the bottom, with the bottom two on the other
side of the node. (i) The superconducting gap (filled circles) at
12K and the pseudogap (open circles) at 49K, as a function of the
Fermi surface angle ($\phi$), with the solid curve the simple {\it
d}-wave form $\Delta_{max}\cos(2\phi)$ with $\Delta_{max}$ = 25 meV.
The squares are the values of the inverse lifetime, $\Gamma_1$, at
12K, which also follows a simple {\it d}-wave form (solid curve).
(j) The Fermi surface
at 12K (filled circles) and at 49K (open circles). The solid line is a
tight-binding fit. (k) Scaling of $\Gamma_1$ with $\Delta$ at 12K
with an extrapolation to the node by the dashed line.} \label{fig1}
\end{figure*}

In Fig.~1 we show spectra obtained below and above $T_c$ from angle
resolved photoemission spectroscopy (ARPES) for underdoped
La$_{1.895}$Sr$_{0.105}$CuO$_4$ ($T_c$=30K) near the node
(Fig.~1(a), (d)), near the anti-node (Fig.~1(c), (f)), and in
between (Fig.~1(b), (e)). Here, the node refers to where the {\it
d}-wave superconducting gap vanishes, with the anti-node where
it is maximal. To determine the Fermi momentum $k_F$ and the
energy gap, we symmetrize the spectra along each
cut~\cite{Nat98}, $A(\omega) = I(\omega) + I(-\omega)$,
where $I(\omega)$ is the measured intensity at the energy $\omega$.
$k_F$ was identified by searching for the minimum gap location of
the symmetrized spectra along a given cut in momentum space. The
same underlying Fermi surface was obtained by analyzing the spectra
in the superconducting state and in the pseudogap phase. The energy
gap at a given $k_F$ was determined by fitting the symmetrized
spectrum (Fig.~1(g)), with the spectral function calculated from a
model self-energy~\cite{PRB98} of the form $\Sigma = -i\Gamma_1 +
\Delta^2/(\omega+i0^+)$, and then
convolved with the instrumental resolution. Like for lightly
underdoped La$_{1.855}$Sr$_{0.145}$CuO$_4$~\cite{Ming}, the
superconducting gap at $T$ = 12 K traces out a simple {\it d}-wave
form with a maximum amplitude of 25 meV at the anti-node
(Fig.~1(i)). Above $T_c$ at $T$ = 49K, there is a small gapless
Fermi arc near the node (Fig.~1(h), (i)). Beyond the Fermi arc, a
pseudogap is present that follows the same simple {\it d}-wave form
as the superconducting gap. It is interesting to note that in our
spectral fits, the inverse lifetime, $\Gamma_1$, has the same
angular anisotropy as the energy gap (Fig.~1(i), (k)). Similar
results have been inferred from fits to scanning tunneling
microscopy data below $T_c$~\cite{Alldredge}, and
photoemission data, both below $T_c$~\cite{Chang} and also
above $T_c$ in the pseudogap phase~\cite{Adam}.

\begin{figure*}
\includegraphics
[width=5in]
{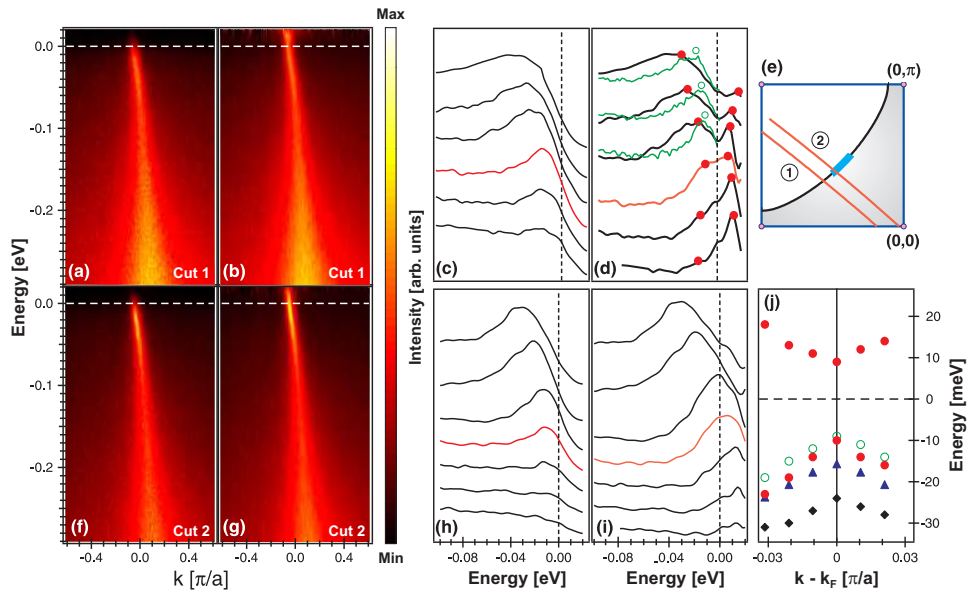}
 \caption{(Color online) ARPES spectra for underdoped La$_{1.895}$Sr$_{0.105}$CuO$_4$
at 49K. (a)-(d): the spectra along cut 1 in (e). (a) the ARPES
intensity, (b) the intensity divided by the resolution broadened
Fermi function, (c) the spectra in the vicinity of $k_F$, and (d)
the Fermi function divided spectra. The thick and thin lines in (d)
are the spectra at 49K and 12K, respectively. The closed and open
circles indicate the spectral peak positions. (e) Fermi surface and
locations of cuts 1 and 2. The thick line centered at the node
indicates the gapless Fermi arc. (f)-(i): the same as (a)-(d), but
the spectra are along cut 2 in (e). (j) The dispersions in the
gapped region of the zone obtained from the Fermi function divided
spectra. The pair of closed circles are the two branches of the
dispersion derived from (d) at 49K, the dispersion indicated with
open circles is along the same cut (cut 1 in (e)) but at 12K. The
curves indicated by triangles and diamonds are the dispersions at
49K in the vicinity of the lower underlying Fermi surface along cut
2 and 3 in Fig. 1(j), respectively.} \label{fig2}
\end{figure*}

Fig.~2 demonstrates the different dispersions in the pseudogap phase
along two momentum cuts, one where the spectrum is gapped (cut 1 in
Fig.~2(e)) and the other which intersects the gapless Fermi arc (cut
2 in Fig.~2(e)).  As the spectra were acquired at $T$ = 49K, there
is appreciable thermal population above the Fermi energy ($E_F$),
which allows us to follow the dispersions through $E_F$. To trace
the dispersion in the vicinity of $E_F$ (Fig.~2(j)), we divide the
raw ARPES spectra by the instrumental resolution broadened Fermi
function, and then follow the peak positions of the divided spectra.
When the spectral peak is weak and sits on a sloped background (for
the lowest two curves in Fig.~2(d)), we first fit the spectrum with
polynomials to high precision and then use the second derivative of
the fitted curve to determine the peak position. Along cut 2, the
spectral peak of the Fermi function divided data continuously moves
to higher energies and crosses $E_F$ at the underlying $k_F$
(Fig.~2(i)). The dispersion resembles that of a normal metal.  On
the other hand, the dispersion along cut 1 shows a remarkable
difference to that along cut 2. As shown in Fig.~2(c), from top to
bottom curves, the spectral peak approaches $E_F$ before $k$ reaches
the underlying $k_F$ and then it recedes to higher binding energies
and loses spectral weight. Fig.~2(d) shows the Fermi function
divided data along cut 1. The dispersion along this cut has two
branches, separated by an energy gap, moving in opposite directions:
one is above $E_F$ and the other is below. As $k$ approaches the
underlying $k_F$ from the occupied side, the spectral weight of the
upper branch increases. The spectral peaks of the two branches have
approximately the same weight at $k_F$, and at the same location,
the energy gap between the two branches is minimal, giving rise to a
flat topped spectrum. All of this is characteristic of the
Bogoliubov-like dispersion seen previously below
$T_c$~\cite{JC-PRB96,Matsui}. But, away from $k_F$, the
peaks in the lower and upper branches are slightly asymmetric in
energy relative to $E_F$ (Fig.~2(j)). This may result from an
asymmetry of the self-energy in the pseudogap phase, which would act
to broaden the spectral peaks and shift them to slightly higher
binding energy on the occupied side. To illustrate this, in
Fig.~2(d), we overlay the spectra obtained in the superconducting
state (the thin lines) on those in the pseudogap phase. It can be
seen that the peak positions of the Fermi function divided spectra
in the superconducting state are at a lower binding energy, and the
dispersion is now symmetric relative to $E_F$ to that of the upper
branch determined in the pseudogap phase. It should be mentioned
that in the gapped region of the zone in the pseudogap phase, all
dispersions obtained from the Fermi function divided spectra show a
back bending at $k_F$ (Fig.~2(j)), one of the characteristics of the
Bogoliubov-like dispersion. However, when the energy gap amplitude
becomes too large near the anti-node, only the lower branch of the
dispersion can be measured. Similar back bent dispersions have been
reported for underdoped Bi2212 above $T_c$~\cite{Kanigel}.
We note that the spectral peaks above $E_F$ are narrower than those
below - this may be due to the asymmetry effect we noted above, or
could be an artifact of the Fermi function division process, which
is not an exact procedure because the observed spectra result from a
convolution with the instrumental resolution function.

\begin{figure*}
\includegraphics
[width=4.5in]
 {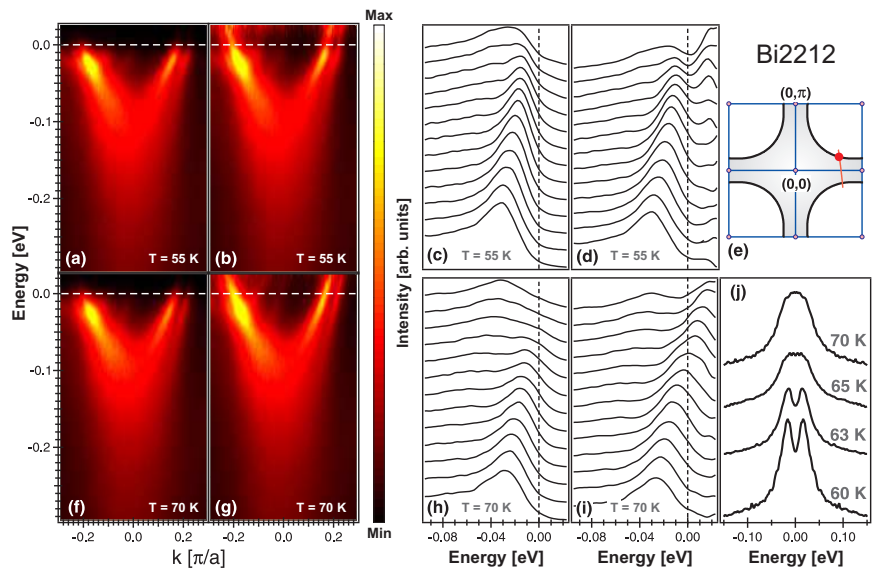}
 \caption{(Color online) ARPES spectra for overdoped
Bi$_2$Sr$_2$CaCu$_2$O$_{8+\delta}$ ($T_c$ = 65K). (a)-(d): the
spectra taken at 55K along the cut in (e). (a) the ARPES intensity,
(b) the intensity divided by the resolution broadened Fermi
function, (c) the spectra in the vicinity of $k_F$ (solid dot in
(e)), and (d) the Fermi function divided spectra. (e) The Fermi
surface and the location of the cut. (f)-(i): the same as (a)-(d),
but the spectra are taken at 70K. (j) the symmetrized spectra at the
anti-node at various temperatures. The superconducting gap closes at
$T_c$.} \label{fig3}
\end{figure*}

The spectral change from cut 1 to cut 2 in Fig.~2(e) is similar to
the spectral change from the superconducting state to the normal
state of overdoped cuprates which do not have a pseudogap above
$T_c$.  To demonstrate this, we present in Fig.~3 ARPES spectra
(also divided by the resolution broadened Fermi function) along the
momentum cut shown in Fig.~3(e) on a heavily overdoped
Bi$_2$Sr$_2$CaCu$_2$O$_{8+\delta}$ (Bi2212) sample with $T_c$ = 65K
and a maximum superconducting gap, $\Delta_{max}$= 24 meV, similar
to that of underdoped La$_{1.895}$Sr$_{0.105}$CuO$_4$.  Above $T_c$,
the spectral peak of the Fermi function divided spectra shows a
linear dispersion going through $k_F$ (Fig.~3(g), (i)). Below $T_c$
(Fig.~3(b), (d)), the superconducting gap opens up and the linear
dispersion of the normal state transforms into the Bogoliubov
dispersion of the superconducting state:
$E_{k}=\pm\sqrt{\epsilon_{k}^{2}+\Delta_{k}^{2}}$, where
$\epsilon_k$ is the energy band dispersion in the normal state and
$\Delta_k$ is the gap function. The double branches in the
electronic excitation spectra involve mixtures of electron and hole
states, the spectral weight in the lower (upper) branch is
proportional to $v^2$ ($u^2$) through the relation
$u^{2}=1-v^{2}=\frac{1}{2}(1+\epsilon_{k}/E_{k})$, where $u$ and $v$
are the coherence factors~\cite{JC-PRB96,Matsui}. At
the normal state $k_F$, the lower (upper) branch of the Bogoliubov
dispersion reaches its maximum (minimum), and the two branches have
equal spectral weight. Remarkably, like in overdoped Bi2212, the two
branches of the dispersion along cut 1 in the pseudogap phase of
underdoped La$_{1.895}$Sr$_{0.105}$CuO$_4$ possess all of these
properties.

To summarize, our main experimental findings are: 1) beyond the
gapless Fermi arc, the pseudogap above $T_c$ has the same simple
{\it d}-wave form as the superconducting gap below $T_c$, 2) above
{\it T}$_c$ there exists a Bogoliubov-like dispersion near the
underlying $k_F$ where the spectra are gapped, and 3) the same
underlying Fermi surface was obtained both in the superconducting
state and in the pseudogap phase. In the pseudogap phase, the low
energy electronic excitations along this cut are well defined both
in energy and in momentum (Fig.~2(a)-(d)). Thus our experimental
results can readily rule out the possibility that the pseudogap in
underdoped cuprates only exists in the anti-nodal region, which has
been attributed to a competing order such as charge
ordering~\cite{KMShen}. Our experimental findings are also
inconsistent with a wide range of scenarios where the pseudogap
originates from some ordering phenomenon associated with a non-zero
$Q$ vector~\cite{mikePRB07}. In our ARPES
spectra, we do not find any signature of additional (shadow) states
that are displaced by a non-zero $Q$, neither in the pseudogap phase
nor in the superconducting state. It should be mentioned that in the
pseudogap phase, the dispersion in the anti-nodal region also bends
back at the underlying $k_F$ (Fig.~2(j)). However, due to the large
amplitude of the pseudogap near the anti-node, the upper branch of
the Bogoliubov-like dispersion is not thermally populated at 49K,
and thus cannot be identified by Fermi function division.

Our experimental results support the idea that the pseudogap
originates from preformed Cooper pairs for $T > T_{c}$.
However, because the energy gap is larger than the phase stiffness
in the pseudogap phase, the preformed pairs have a finite lifetime
and can not travel in the crystal coherently. The similarity between the
Bogoliubov-like dispersion in the
pseudogap phase of underdoped La$_{1.895}$Sr$_{0.105}$CuO$_4$
and the dispersion of Bogoliubov quasiparticles in the
superconducting state of heavily overdoped Bi2212 provides direct
evidence that the pseudogap in underdoped cuprates arises from
pairing of electrons, and thus the pseudogap phase is a state precursor to
superconductivity. Because the superconducting and pseudogaps have
the same simple {\it d}-wave form along the
underlying Fermi surface beyond the Fermi arc (Fig.~1(i)), it is difficult to
image that there is  some $k_F$ located between the
anti-node and the end of the Fermi arc at which the pseudogap
changes its nature from preformed
Cooper pairs near the arc to competing order near the anti-node.
 It is more reasonable to infer that there is a single pseudogap above $T_c$ which
transforms into the superconducting gap below $T_{c}$.

This work was supported by the Swiss National Science Foundation
(through NCCR, MaNEP, and grant Nr 200020-105151), the Ministry of
Education and Science of Japan, the Swedish Research Council, the
U.S. DOE, Office of Science, under Contract No.~DE-AC02-06CH11357,
and by NSF DMR-0606255. We thank the beamline staff of X09LA at the
SLS for their excellent support.


\begin{thebibliography}{99}


\bibitem{Loeser}
A. G. Loeser \etal, Science {\bf 273}, 325 (1996).

\bibitem{Ding}
H. Ding \etal, Nature {\bf 382}, 51 (1996).

\bibitem{NPK}
M. R. Norman, D. Pines and C. Kallin, Adv. Phys. {\bf 54}, 715
(2005).

\bibitem{Randeria}
M. Randeria \etal, Phys. Rev. Lett. {\bf 69}, 2001 (1992).

\bibitem{Emery}
V. J. Emery and S. A. Kivelson, Nature {\bf 374}, 434 (1995).

\bibitem{Tanaka}
K. Tanaka \etal, Science {\bf 314}, 1910 (2006).

\bibitem{Kondo}
T. Kondo \etal, Phys. Rev. Lett. {\bf 98}, 267004 (2007).

\bibitem{Terashima}
K. Terashima \etal, Phys. Rev. Lett. {\bf 99}, 017003 (2007).

\bibitem{Lee}
W. S. Lee \etal, Nature {\bf 450}, 81 (2007).

\bibitem{Mansson}
M. M\aa{}nsson \etal, Rev. Sci. Instr. {\bf 78}, 076103 (2007).

\bibitem{Nat98}
M. R. Norman \etal, Nature {\bf 392}, 157 (1998).

\bibitem{PRB98}
M. R. Norman \etal, Phys. Rev. B {\bf 57}, 11093 (1998).

\bibitem{Ming}
M. Shi \etal, Phys. Rev. Lett. {\bf 101}, 047002 (2008).

\bibitem{Alldredge}
J. W. Alldredge \etal, Nat. Phys. {\bf 4}, 319 (2008).

\bibitem{Chang}
J. Chang \etal, arXiv:0708.2782.

\bibitem{Adam}
A. Kaminski \etal, Phys. Rev. B {\bf 71}, 014517 (2005).

\bibitem{JC-PRB96}
J. C. Campuzano \etal, Phys. Rev. B {\bf 53}, 14737 (1996).

\bibitem{Matsui}
H. Matsui \etal, Phys. Rev. Lett. {\bf 90}, 217002 (2003).

\bibitem{Kanigel}
A. Kanigel \etal, Phys. Rev. Lett. {\bf 101}, 137002 (2008).

\bibitem{KMShen}
K. M. Shen \etal, Science {\bf 307}, 901 (2005).

\bibitem{mikePRB07}
M. R. Norman \etal, Phys. Rev. B {\bf 76}, 174501 (2007).

\end{thebibliography}
\end{document}